\documentclass[12pt,a4]{article}
\usepackage{epsfig}
\usepackage{amsfonts}
\title {\sc A Non-linear Dynamical Systems' Proof of \\Kraft-McMillan Inequality and its Converse}
\author{Nithin Nagaraj \\ School of Natural  Sciences and Engineering \\ National Institute of Advanced Studies \\ {\bf \small Email:~nithin\_nagaraj@yahoo.com}}
\date{October 31, 2007}

\begin{document}
\maketitle \noindent In this short paper, we shall provide a
dynamical systems' proof of the famous Kraft-McMillan inequality and
its converse. Kraft-McMillan inequality is a basic result in
information theory which gives a necessary and sufficient condition
for the lengths of the codewords of a code to be uniquely
decodable~\cite{Kraft49,McMillan56,Karush61}.

\section{Kraft-McMillan Inequality} \noindent Given a
binary prefix code set $C$ for an alphabet set $A$, the codewords
$c_1, c_2, \ldots, c_N$ with lengths $l_1, l_2, \ldots, \l_N$
necessarily satisfy:

\begin{equation}
\sum_{i=1}^{N} 2^{-l_i} \leq 1
   \label{eqn:eqn1}
\end{equation}

where $N=|A|$, the cardinality of set $A$. A  binary prefix code $C$
is a set of binary codewords such that no codeword is a prefix of
another. Prefix codes are known to be uniquely decodable and easy to
decode. A famous example of prefix codes are the Huffman codes which
have minimum redundancy.

\subsection*{The Binary map} 
\begin{figure}[h]
\centering
\includegraphics[scale=.5]{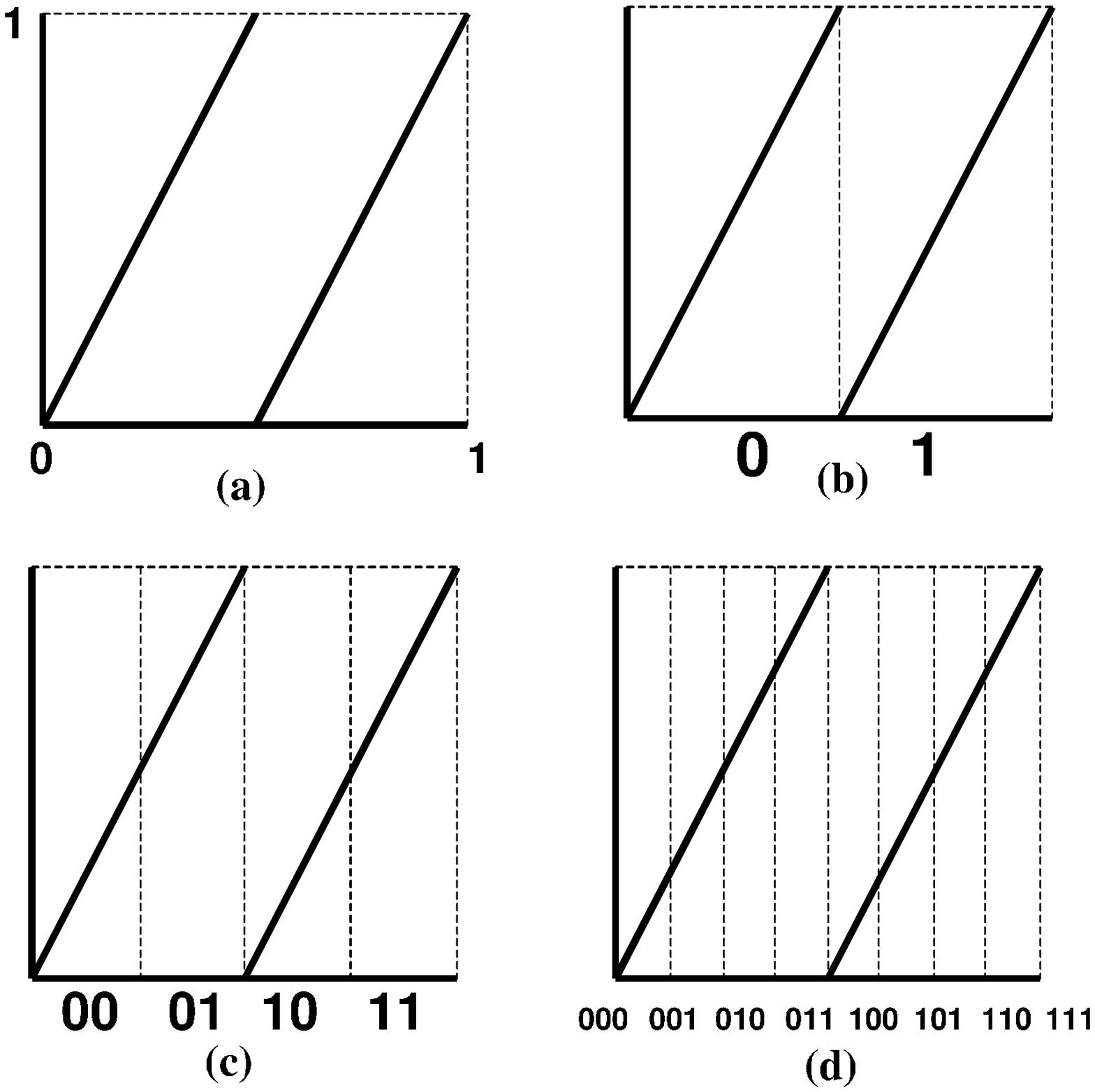}
\caption{(a) Binary Map. (b) Symbolic sequences of length 1. (c)
Symbolic sequences of length 2. (d) Symbolic sequences of length 3.
} \label{fig:figbin1}
\end{figure}

Consider the binary map
(Fig.~\ref{fig:figbin1}) $T:[0,1) \rightarrow [0,1)$:
\begin{eqnarray*}
x & \mapsto & 2x,~~~~~~~~~~~~~~~~~ 0 \leq x<\frac{1}{2}\\
    & \mapsto & 2x-1,~~~~~~~~~~~~ \frac{1}{2} \leq x < 1.
\end{eqnarray*}

It is well known that the binary map is a non-linear chaotic
dynamical system, which preserves the Lebesgue measure (ordinary
length measure). We shall prove two simple lemmas regarding the
binary map which will be used to prove the Kraft-McMillan's
inequality.

\subsection*{{\it Lemma 1:}} Given any sequence (or string) $S$ of 0s and 1s of length $m$, there exists
an unique interval on the binary map of length $2^{-m}$ such that
all initial conditions in that interval will have $S$ as the binary
symbolic sequence corresponding to the first $m$ iterations.

\subsubsection*{{\it Proof:}} Consider the given string $S$ of length $m$ as
a binary prefix in $[0,1)$ (i.e. think of $S$ as $0.S$ in binary).
The interval $[0.S\overline{0}, 0.S\overline{1})$, where the
overline indicates infinite repetition, consists of all possible
binary numbers in $[0,1)$ which have $S$ as the desired prefix. All
these binary numbers when fed as initial conditions to the binary
map will yield $S$ as the symbolic sequence in $m$ iterations (this
is because the binary map can be thought of as a shift map which
spits out the leading bits of the binary representation of the
initial condition). The length of this interval is $0.S\overline{1}
- 0.S\overline{0}$ which is $2^{-m}$.$\hfill \square$

\subsection*{{\it Lemma 2:}}
Two symbolic sequences $S_1$ and $S_2$ which are not prefixes of
each other correspond to two {\bf disjoint} intervals of lengths
$2^{-m_1}$ and $2^{-m_2}$ respectively, where $m_1$ and $m_2$ are
the lengths of $S_1$ and $S_2$ respectively.

\subsubsection*{{\it Proof:}}
The proof is obvious.$\hfill \square$
\section*{{\it Proof of Kraft-McMillan Inequality}} Since $c_1, c_2, \ldots, c_N$ with lengths
$l_1, l_2, \ldots, \l_N$ are prefix codes, using Lemma 1 and 2,
these can be seen as symbolic sequences of {\bf disjoint} intervals
on [0,1) with lengths $2^{-l_1}, 2^{-l_2}, \ldots, 2^{-l_N}$
respectively. Any collection of disjoint intervals on [0,1)
necessarily satisfy Equation~\ref{eqn:eqn1}.$\hfill \square$
\section{Converse of Kraft-McMillan inequality}
\noindent Given a set of codeword lengths that satisfy
Equation~\ref{eqn:eqn1}, there exists a uniquely decodable binary
prefix code with these codeword lengths.

\subsection*{{\it Proof:}}
Let $l_1, l_2, \ldots, \l_M$ be the specified {\bf distinct}
codeword lengths such that they satisfy Equation~\ref{eqn:eqn1}.
Without loss of generality, let us assume that $l_1 < l_2 < \ldots <
\l_M$. Let there be $a_1$ codewords of length $l_1$, $a_2$ codewords
of length $l_2$ and so on up to $a_M$ codewords of length $l_M$.
Kraft-McMillan inequality can be re-written as:

\begin{equation}
\sum_{i=1}^{M} a_i 2^{-l_i} \leq 1, ~~~~~ \sum_{i=1}^{M} a_i = N.
   \label{eqn:eqn2}
\end{equation}
where  $N=|A|$ as before. Let us determine the maximum number of
codewords that can have a particular codeword length $l_i$ while
still satisfying Equation~\ref{eqn:eqn2}. If there are $2^{l_i}+1$
or more codewords with length $l_i$, then $(2^{l_i}+1)2^{-l_i} = 1 +
2^{-l_i} > 1$ violating Equation~\ref{eqn:eqn2}. Thus there can at
most be $2^{l_i}$ codewords of length $l_i$.
\begin{figure}[h]
\centering
\includegraphics[scale=.5]{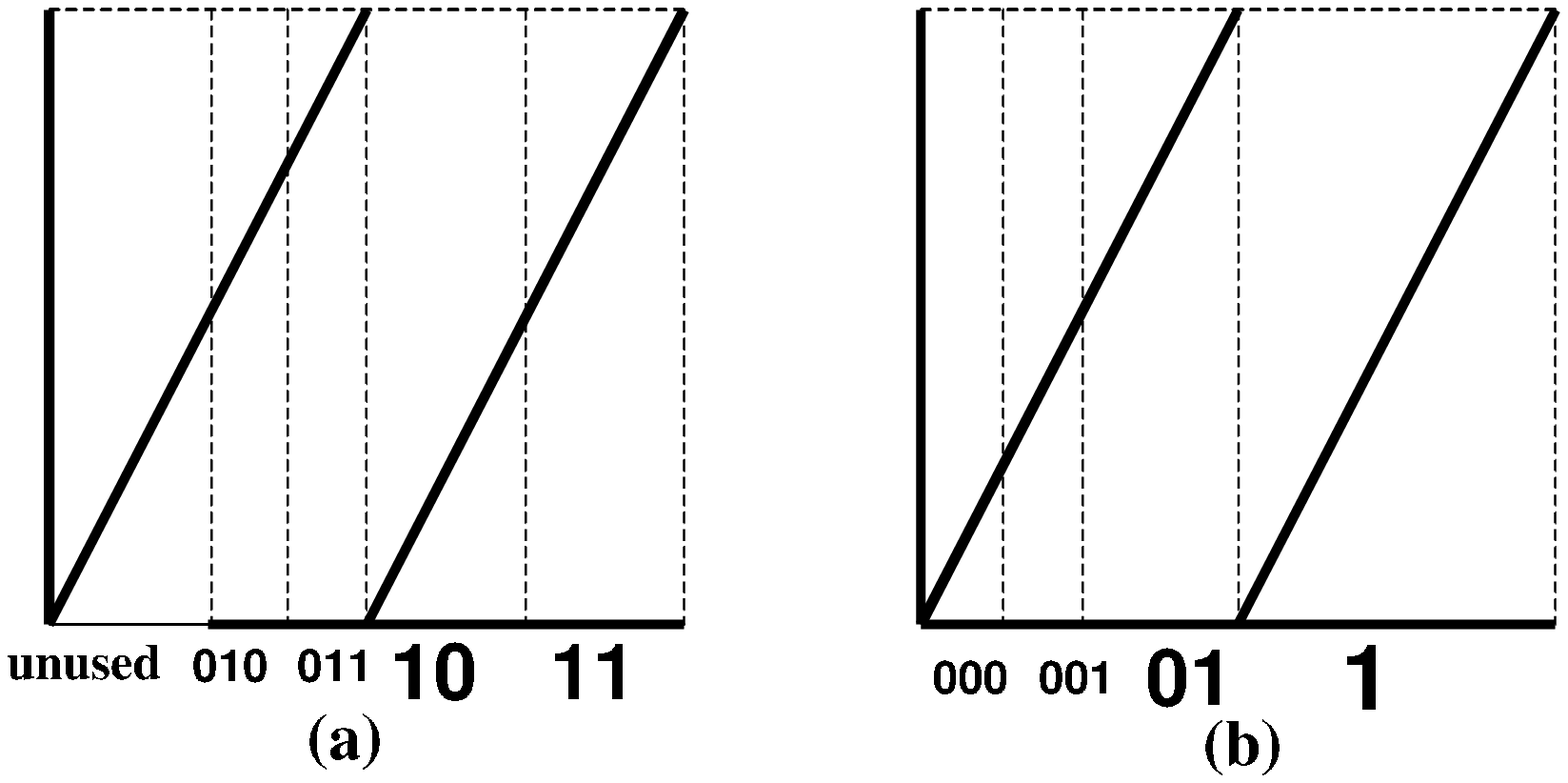}
\caption{Converse of Kraft-McMillan inequality. (a) Assigning
intervals to codewords of lengths \{3, 3, 2, 2\} which satisfies the
Kraft-McMillan inequality. (b) Assigning intervals to codewords of
lengths \{3, 3, 2, 1\} which satisfies Kraft-McMillan inequality
with equality. This is known as a {\it complete} code.}
\label{fig:figbin2}
\end{figure}

Let us begin with $l_1$. We know that there are exactly $2^{l_1}$
disjoint intervals with length $2^{-l_1}$ on the binary map which
have symbolic sequence of length $l_1$. Since the intervals are
disjoint, the symbolic sequences are necessarily prefix codewords.
We first assign the symbolic sequences of $a_1$ of these disjoint
intervals as codewords. Once $a_1$ disjoint intervals of length
$2^{-l_1}$ were used up, we have lost $a_1 2^{l_2-l_1}$ intervals of
length $2^{-l_2}$. The number of available disjoint intervals of
length $2^{-l_2}$ is $2^{l_2} - a_1 2^{l_2-l_1}$. If $a_2 < 2^{l_2}
- a_1 2^{l_2-l_1}$ then we can allocate disjoint intervals to $a_2$
codewords of length $l_2$. This requires $a_2 < 2^{l_2} (1-a_1
2^{-l_1})$, which reduces to $a_1 2^{-l_1} + a_2 2^{-l_2} < 1$ which
is necessarily true from Equation~\ref{eqn:eqn2}. Thus we can use
the symbolic sequence of $a_2$ disjoint intervals as prefix
codewords (of length $l_2$). This argument is repeated for $a_3$ and
so on until we have allocated unique disjoint intervals to all
codewords (see example in Fig.~\ref{fig:figbin2}). We have thus
proved the converse of Kraft-McMillan inequality by construction of
prefix codewords using symbolic sequences of disjoint intervals on
the binary map.$\hfill \square$

The arguments above can be extended in a straightforward manner for
ternary and higher bases. In the case of codewords of base-$B$, the
$B$-ary dynamical system is used ($x \mapsto Bx~mod~1$ for all $x
\in [0,1)$).

\end{document}